\documentclass[prb,onecolumn,nofootinbib,citeautoscript,10pt,longbibliography,notitlepage]{revtex4-2}

\usepackage{graphicx}
\usepackage{dcolumn}
\usepackage{bm}
\usepackage{color} 
\usepackage{amsmath}
\usepackage{tabularx,graphicx}
\usepackage{epstopdf}
\usepackage{latexsym}
\usepackage{amssymb}
\usepackage{amsmath}
\usepackage{color, colortbl}
\usepackage{psfrag}
\usepackage{bbm}
\usepackage{bm}
\usepackage{titlesec}
\usepackage{dsfont}
\usepackage{feynmp}
\usepackage{slashed}
\usepackage{multirow}
\usepackage[tight]{subfigure}

\usepackage[papersize={8.5in,11in}]{geometry}

\usepackage{color}
\definecolor{darkblue}{rgb}{0.,0.,0.4}
\definecolor{darkred}{rgb}{0.5,0.,0.}
\definecolor{BlueViolet}{RGB}{138,43,226}
\definecolor{SkyBlue}{RGB}{30,144,255}
\definecolor{DarkGreen}{RGB}{0,100,0}
\usepackage[pdftex,colorlinks=true,linkcolor=darkblue,citecolor=blue,urlcolor=darkred]{hyperref}

\usepackage{physics}

\renewcommand{\epsilon}{\varepsilon}

\def \nn{\nonumber \\}

\begin{document}

\title{Identifying gap-closings in open non-Hermitian systems by Biorthogonal Polarization}

\author{Ipsita Mandal}
\email{ipsita.mandal@snu.edu.in}

\affiliation{Department of Physics, Shiv Nadar Institution of Eminence (SNIoE), Gautam Buddha Nagar, Uttar Pradesh 201314, India
\\and\\
Freiburg Institute for Advanced Studies (FRIAS), University of Freiburg, D-79104 Freiburg, Germany}

\begin{abstract}
We investigate gap-closings in one- and two-dimensional tight-binding models with two bands, containing non-Hermitian hopping terms, and open boundary conditions (OBCs) imposed along one direction. We compare the bulk OBC spectra with the periodic boundary condition (PBC) spectra, pointing out that they do not coincide, which is an intrinsic characteristic of non-Hermitian systems. The non-Hermiticity, thus, results in the failure of the familiar notions of bulk-boundary correspondence found for Hermitian systems. This necessitates the search for topological invariants which can characterize gap-closings in open non-Hermitian systems correctly and unambiguously. We elucidate the behaviour of two possible candidates applicable for one-dimensional slices --- (1) the sum of winding numbers for the two bands defined on a generalized Brillouin zone and (2) the biorthogonal polarization (BP). While the former shows jumps/discontinuities for some of the non-Hermitian systems studied here, at points when an edge mode enters the bulk states and becomes delocalized, it does not maintain quantized values in a given topological phase. On the contrary, BP shows jumps at phase transitions, and takes the quantized value of one or zero, which corresponds to whether an actual edge mode exists or whether that mode is delocalized and absorbed within the bulk (not being an edge mode anymore).
\end{abstract}

\maketitle


\section{Introduction}

The study of topological phases in non-Hermitian systems has taken a centre stage in mainstream condensed matter physics ever since it has been realised that stable band-crossing points are more generic and abundant than the Hermitian counterparts \cite{emil_nh_nodal,emil_review,ips-emil,prl_eps}. Stable nodal phases in \textit{non-Hermitian Hamiltonians} involve the emergence of \textit{Exceptional Points} (EPs), which are singular points at which two or more eigenvalues, along with their eigenvectors, coalesce \cite{Berry2004,Heiss_2012,PhysRevX.6.021007,ep-optics,ozdemir2019parity}. The EPs are intimately connected to topological phase transitions in generic contexts \cite{ips-ep-epl,ips-tewari,emil_review,kang-emil,ips-emil,prl_eps,ips-kang,ips-yao-lee}. In addition to EPs, non-Hermitian tight-binding lattice models exhibit unusual localization phenomena like the non-Hermitian skin effect \cite{hatano-nelson,hatano-nelson2, ssh_lee, yaowang, slager, Okuma2020, Martinez2018, Xiao2020, okuma2023, Lin2023, kang-emil, maria_emil, ips-yao-lee}. In this paper, we continue the ongoing efforts to characterize non-Hermitian topological phase transitions, as they exhibit fundamentally distinct properties compared to the well-understood Hermitian topological phases, which we outline below. 

Non-Hermitian Hamiltonians exhibit novel topological phases because of their complex eigenvalues \cite{PhysRevB.101.205417,mong}. For a tight-binding lattice model with periodic boundary conditions (PBCs), the spectrum can be straightforwardly obtained by Fourier transforming to the momentum space. The components of the momentum vector form a $d$-dimensional torus, which is the Brillouin zone (BZ) in the reciprocal space. Nontrivial topological phases appear when the eigenvectors and eigenvalues are twisted under these toric boundary conditions. The twists of the eigenvectors, familiar in the Hermitian settings, can be described by topology of the vector bundles formed by the eigenstates. In this context, one good example is the two-dimensional (2d) Chern insulator. The Chern class of the eigenvector bundle can be expressed as the Berry curvature, whose integral over the BZ gives the Chern number. However, when complex eigenvalues are present, the twists of the eigenvectors can no longer uniquely describe the topology of the system. The twists of the eigenvalues can described by a new topological invariant which is the braid of energy spectra and the Chern number associated with the eigenvectors reduces to a fragile topological invariant \cite{mong}.

\begin{figure}[t]
    \centering
    \includegraphics[width=0.9\linewidth]{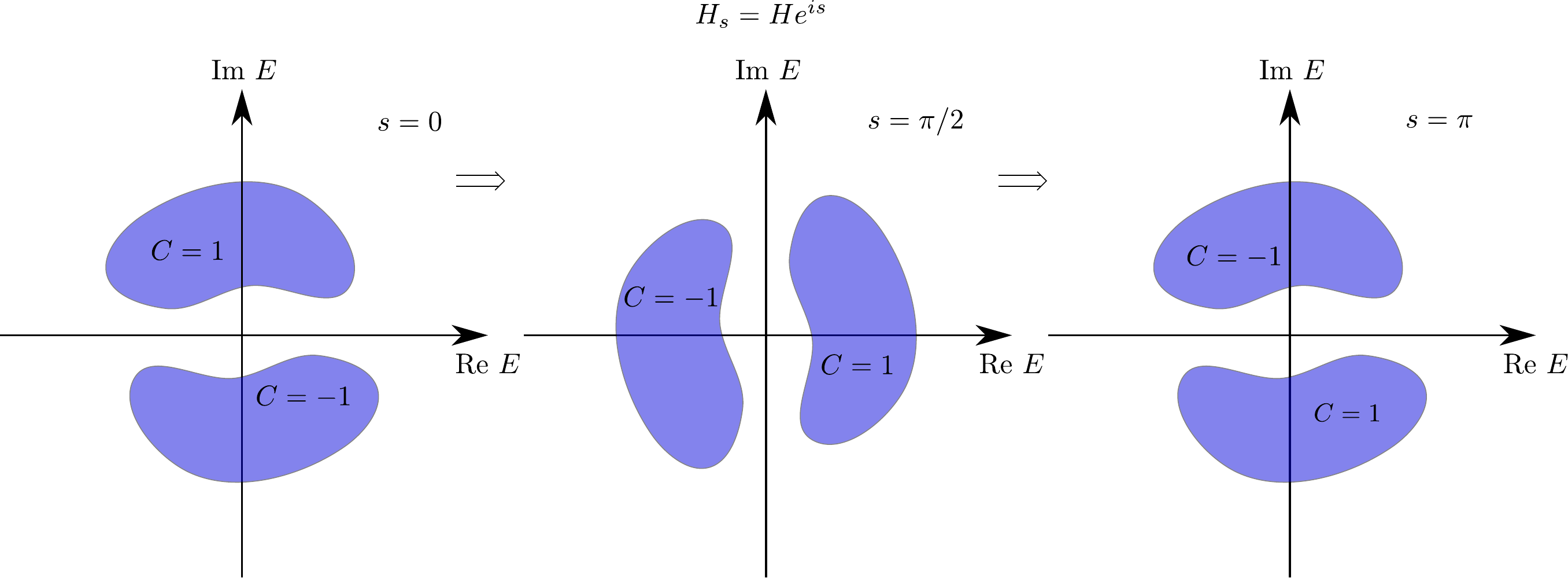}
    \caption{For non-Hermitian Hamiltonians, Chern numbers differing by a sign cannot be distinguished topologically. This follows from the fact that the upper and the lower bands can be interchanged continuously without closing the gap between them in the complex plane. This means that the phase transition between $C=1$ and $C=-1$ states, allowed for a Hermitian Hamiltonian, can be avoided here in an enlarged parameter space made possible by the presence of complex parameter(s).}
    \label{fig_crdc}
\end{figure}

In order to review the idea of braid topology, let us first look at the one-dimensional (1d) cases, where the BZ is equivalent to a circle, labelled by the momentum $k$. In Hermitian scenarios, there are no topological phases in the absence of a symmetry, as the eigenvectors cannot be twisted along a circle. However, in non-Hermitian systems, the phases can be nontrivial even in the absence of symmetry. The eigenvalues of the system can be permuted when traversing the 1d BZ. Assuming we have $p$ eigenvalues labelled as $(E_1,E_2,\ldots, E_p)$ at $k=0$, as the momentum increases, these eigenvalues can move around each other. When $k$ gradually increases to $k=2\, \pi$, the eigenvalues may not go back to themselves, but rather take the form $(E_{i_1},E_{i_2},\ldots, E_{i_p})$, with $(i_1,i_2,\ldots, i_p)$ being a permutation of $(1,2,\dots, p)$.  Such topology is described by the braid group $B_p$, where $p$ is the number of bands/orbitals. When $p=2$, the braid group is Abelian $B_2=\mathbb Z$ and is reduced to the energy vorticity \cite{shenzhenfu,leykambliokhhuangchongnori}.

In two dimensions, since the BZ is a torus, there are two natural nontrivial circles on a torus, the meridian and the longitude. Without loss of generality, we can parametrize them to be along $k_x$ and $k_y$, respectively. The spectrum of a gapped Hamiltonian can have nontrivial braids of energy along these two circles and the topological phases are characterized by two elements of the braid group. On top of the eigenvalue topology, the eigenvectors can also have nontrivial twists. For two-band models, the eigenvector topology has been classified in Refs.~\cite{PhysRevB.101.205417,mong}. It can still be described by an integer, the Chern number. However, not every Chern number describes distinct phases in non-Hermitian models --- different Chern numbers may belong to the same topological phase. They are manifested as an $\mathbb N$ (natural number) invariant or a $\mathbb Z_2$ invariant (see Table~\ref{tb_tpph}). 
Therefore, a novel feature of two-band non-Hermitian Chern insulators is that the signs of the Chern number no longer distinguish different topological phases. This follows from the fact that there is no ordering of eigenvalues. We demonstrate this with the help of a simple example when there is no braid and the two bands are well separated such that the energy of the two bands form two disconnected sets in the complex plane. Without any loss of generality, we take them to be symmetric with respect to $E=0$. The Chern numbers $C $ and $C'$ of the two bands are equal and opposite, resulting in $C+C'=0$. However, unlike Hermitian scenarios, there is no canonical identification of the lower band. In fact, we can always adiabatically swap the two bands, without closing the gap in-between, by the transformation $H\to H\exp(i \,s)$, where $s$ can go from $0$ to $\pi$ (cf. Fig.~\ref{fig_crdc}). The Chern number of the band in the lower half plane switches sign during this adiabatic deformation. Consequently, there is no topological distinction between phases characterized by $C$ and $- \, C$.

The above discussions involve PBCs along all directions of a given lattice. Things get even more complicated when we impose open boundary conditions (OBCs) along one (or more) directions, depending on the dimensionality of the lattice under consideration. A crucial aspect in the study of non-Hermitian Hamiltonians is the fact that the PBC spectra and OBC spectra are completely different, with a complete breakdown of the familiar bulk-boundary correspondence of Hermitian situations \cite{slager,Okuma2020,emil_review}. 
In this paper, we consider the question of identifying gap-closing points in 1d and 2d non-Hermitian Hamiltonians, featuring two bands, with the help of a topological invariant. We consider OBCs along one direction, which gives us slices of finite 1d chains. We start with the conventional quantities like the winding number defined over the so-called \textit{generalized Brillouin zone} \cite{yaowang,yokomizo}, which involves extending the Bloch band theory to non-Hermitian systems with open boundaries~\cite{yinjiangliluchen,yaowang,yokomizo}. We show that, although in some specific cases, this quantity captures phase transitions by showing jumps in its values when an edge mode enters/leaves the bulk states, it does not retain a uniform quantized value within a given topological phase. Moreover, for some of the systems studied here, it remains zero throughout [cf. Fig.~\ref{fig_cherny}]. Next, we consider another feasible candidate, known as the \textit{biorthogonal polarization} (BP), introduced in Refs.~\cite{flore-elisabet, flore-elisabet0}, whose value takes an exact value of zero or one depending on whether a mode exists as edge mode or is merged with the bulk states. We elucidate our findings for non-Hermitian generalizations of the 1d Su-Schrieffer-Heeger (SSH) model \cite{ssh0, ssh, yaowang,song_yao_ssh, han_ssh,  ssh_lee, yinjiangliluchen, maria_emil} and and the 2d Rice-Mele model \cite{ricemele} (and its variations). Considering the evidence from the systems we study, we conclude that BP is an unambiguous topological invariant for 1d slices (with open boundaries) of non-Hermitian systems.

The SSH model \cite{ssh0, ssh} is a one-dimensional bipartite lattice model that harbours topologically distinct phases.
Although the model was originally proposed to describe solitons in a polyacetylene chain \cite{ssh0, ssh}, it has been realized experimentally in myriad quantum systems like cold-atoms in optical lattices \cite{ssh_expt} and monolayer of chlorine atoms on copper surface \cite{ssh_expt2}. For the SSH chain with PBCs at both ends (which is, of course, Hermitian), the winding number serves as a topological invariant which takes the values zero or one corresponding to a topologically trivial or nontrivial phase, determined by the ratio of the two hopping coefficients \cite{schnyder_review}. The bulk-boundary correspondence then ensures that these two phases have zero or one edge state at each end in the corresponding open chain \cite{schnyder_review, ref_demand}. Since the system is exactly solvable and, thus, easy to analyze, it is one of the most widely used Hamiltonians to demonstrate and explain topological phenomena. One line of generalizations of the original SSH model is the introduction of non-Hermiticity, for example, by introducing a non-reciprocal/asymmetric hopping term \cite{ssh_lee, yaowang, flore-elisabet, flore-elisabet0}. Analogous to the paradigmatic Hermitian SSH model, such non-Hermitian versions have emerged as an important model for investigating non-Hermitian topological phenomena. Hence, we choose to demonstrate our findings for 1d systems using such a non-Hermitian system.

The SSH Hamiltonian augmented by a staggered onsite potential at the sublattice sites (which we denote here by $A$ or $B$, following the general convention) is known as the Rice-Mele model \cite{ricemele}, and is a paradigmatic model of 1d semiconductors with broken inversion symmetry. Hence, it constitutes another simple 1d model to study topological properties of Bloch Hamiltonians. It can be realized in set-ups like attractive ultracold fermions in a shaken optical lattice \cite{ricemele_optical}. Again, the non-Hermitian versions of the Rice-Mele model provide us with invaluable insights about counter-intuitive topological phenomena emerging in non-Hermitian systems \cite{maria_emil, flore-elisabet}. Furthermore, one can construct 2d Chern insulators by stacking
chains described by the Rice-Mele Hamiltonian \cite{flore_2d_chern}. Here, we consider such 2d systems augmented by non-Hermitian terms \cite{flore-elisabet}.

\renewcommand{\arraystretch}{1.5}
\begin{table*}[t]
\centering
\begin{tabular}{ |p{3.0cm}|p{5.5cm}|p{6.5cm}|  }
\hline
\multicolumn{3}{|c|}{Topological gapped phases for two-band models} \\
\hline
\centering 1d & \multicolumn{2}{|c|}{2d}\\
\hline
\centering Eigenvalues: $B_2=\mathbb Z$  & Eigenvalues: $(a,b)\in (\mathbb Z,\mathbb Z)$, both even
\newline 
Eigenvectors: $\mathbb N$
&Eigenvalues: $(a,b)\in (\mathbb Z,\mathbb Z)$, at least one odd
\newline 
Eigenvectors: $\mathbb Z_2$
\\
\hline
\end{tabular}
\caption{Topological two-band non-Hermitian phases in 1d and 2d. The phases in 1d are described by the topology of the eigenvectors. In 2d, the phases are described by both the eigenvalues and eigenvectors. The eigenvalue topology is described by two braid elements, $a,b\in B_2=\mathbb Z$, along the two bigger circles of the BZ. The eigenvector topology is described by a reduced Chern number. When both braids are even, only the absolute value of the Chern number is a topological invariant. When at least one of the two braids is odd, only the mod $2$ value of the Chern number is topologically invariant.}
\label{tb_tpph}
\end{table*}

The paper is organized as follows. In Sec.~\ref{secbp}, we identify various candidates for obtaining topological invariants to identity phase transitions in open non-Hermitian systems and spell out the formalism to compute them. In Sec.~\ref{sec_ssh}, we consider the 1d non-Hermitian SSH model and chalk out the various phases possible at different values of the parameters.  Sec.~\ref{secchern} is devoted to the study of 2d Rice-Mele model with non-Hermitian couplings. From the results in Sec.~\ref{sec_ssh} and Sec.~\ref{secchern}, we conclude that BP emerges as the undisputed topological invariant capturing the gap-closings unambiguously for open 1d chains. Finally, we end with a summary and outlook in Sec.~\ref{secsum}. 

\section{Identifying phase transitions for open boundaries}

\label{secbp}

For generic 1d periodic systems (with states labelled by momentum $k$), the winding number for any separable band with energy $E_n$, can be associated with the non-Hermitian Zak phase (i.e., the Berry phase across the 1d BZ) via the expression
\begin{align}
\nu_{n} =  \frac{1} { 2 \,\pi} 
\oint_{\mathcal{C}} dk \,\langle u_{L,n} |\,i\,\partial_k\,| u_{R,n} \rangle \,,
\end{align}
where $| u_{L,n} \rangle$ and $|u_{R,n} \rangle$ denote the left eigenvector and right eigenvector, respectively, for the given band, satisfying the normalization $\langle u_{L,n} | u_{R,n} \rangle =1$.\footnote{For Hermitian systems, $ |u_{R,n} \rangle = | u_{L,n} \rangle$.}
The integral is taken along a closed loop $ {\mathcal{C}} $ in the momentum space.
In particular, $\nu_{n}$ can have a fractional value in multiples of $1/N$, because the BZ is $2N\pi$ periodic when circling an EP around which the dispersion scales as an $N^{\rm{th}}$ root. For a Hamiltonian without a chiral symmetry, the $\nu_n$'s are not individually quantized, and each can take an arbitrary complex value. Nevertheless, the sum $\nu_{tot}= \sum_n \nu_n $ takes quantized values, which change across the phase boundaries demarcating band-touching points, thus characterizing the various topological phases~\cite{PhysRevA.87.012118}. However, it gives the correct topological phases (and band-touchings) only for a system with chiral symmetry and fails for systems without chiral symmetry \cite{PhysRevA.98.052116}.

For 1d topological non-Hermitian systems with open boundaries, the topological properties can be characterized by the above winding number only after generalizing to non-Bloch band theory \cite{yaowang,yokomizo}. This involves using the concept of the generalized Brillouin zone (GBZ), denoted by the closed contour $ {\mathcal{C}}_{\text{GBZ}}$, by defining $\beta = e^{i\,k}$, where $k $ is generically complex (i.e., not confined to real values). The GBZ is determined by the following procedure: a generalized ``Bloch'' Hamiltonian $H(\beta)$ is obtained for the open system by substituting $k$ by $-i\,\ln \beta $ in the momentum-space Hamiltonian $H(k)$ for the corresponding periodic system:
\begin{align}
\lbrace H(k), \, k \in \mathds{R} \rbrace \rightarrow 
\lbrace H(\beta=e^{i\,k}), \, k \in \mathds{C} \rbrace \,.
\end{align}
The eigenvalue equation $ \left [\beta^p \det[H(\beta) -E]  \right ]=0 $ gives a polynomial equation for $\beta$, where $p$ is the order of the pole of of $\det H$. If the degree of this equation is $2 M$, we need to arrange the roots $\beta_i$ (where $i \in [1,\cdots, 2M]$) in the order
$|\beta_1| \leq |\beta_2|\leq \cdots \leq  |\beta_{2M-1}| \leq |\beta_{2M}|$. From this list, $\Gamma_{\text{GBZ}}$ is obtained from the trajectory of $\beta_{M}$ and $\beta_{M+1}$ under the condition $\beta_{M}=\beta_{M+1}$. The energy spectrum $E$ of the open system is also obtained from these admissible values of the roots.

In this paper, we restrict ourselves to two-band systems of the form 
\begin{align}
\label{h_gen}
H= { \mathbf{d}} ({\mathbf k})  \cdot \boldsymbol\sigma \,.
\end{align}
The eigenvalues for PBC are given by $\pm E_\textrm{PBC} ( {\mathbf k} )$, where
\begin{align}
E_\textrm{PBC} ({\mathbf k})= 
\sqrt{d_x^2 ({\bf k}) + d_y^2 ({\bf k}) +d_z^2 ({\bf k}) }\,.
\end{align}
The corresponding eigenvectors are
\begin{align}
\psi_\pm (\mathbf k) =\Big ( d_z(\mathbf k) \pm E_\textrm{PBC}
\quad  d_x(\mathbf k) + i\,d_y(\mathbf k) \Big )^T
\end{align}
in one and two spatial dimensions.
A Hamiltonian of the form ${ \mathbf{d}} ({\mathbf k})  \cdot \boldsymbol\sigma $ always has the eigenvalues in pairs of $ \pm E$. This means an EP can appear only at $ E = 0$. We consider an open system with OBC imposed along one of the directions.
Defining $H({\mathbf{k}_\perp},\beta) = \mathbf{d} ({\mathbf{k}_\perp},\beta) \cdot \boldsymbol{\sigma}$, where $\beta $-substitution corresponds to the  momentum-component along which OBC is imposed, and $ {\mathbf{k}_\perp}$ denotes the momentum directions along which periodicity is retained.
According to the prescription explained above, the characteristic polynomial for $\beta $ takes the form:
\begin{align}
\label{eq_beta}
\beta^p  \, E^2_\textrm{PBC} ( {\mathbf{k}_\perp}, \beta )
= \beta^p E^2 \,.
\end{align}
Since the admissible roots have the same magnitude, they can be written as $\beta$ and $\beta\, e^{i\,\phi}$ [with $\phi \in [0,2\pi)$]. Substituting these two roots in the above equation, and subtracting the resulting equations from each other, we get the form:
\begin{align}
\label{eq_beta2}
\beta^p\left [
 E^2_\textrm{PBC} ({\mathbf{k}_\perp}, \beta)
 - E^2_\textrm{PBC} ({\mathbf{k}_\perp}, \beta\, e^{i\,\phi})
\right ]=  0 \,.
\end{align}
Here, the $E$-dependence has dropped out, and the equation is easier to solve. Hence, for $\phi$ taking values on the unit circle, we determine the roots $\beta_j$ (with $j \in [1,2M]$) for a given value of $\phi$, and pick out the admissible solutions, which satisfy $\beta_{M}=\beta_{M+1}$, after ordering them in the ascending order according to their absolute values.
For the two-band examples that we consider in this paper, the sum $\nu_{tot}= \sum_n \nu_n $ for the pair of bands shows well-defined jumps at the phase boundaries, which characterize gap-closings of the complex energy bands, for some of the systems studied here (for example, as illustrated in Figs.~\ref{fig_ssh} and \ref{fig_chernx}). 
However, $\nu_{tot}$ deviates from exact quantized values within a given phase [for example, compare subfigures (a) and (b) with subfigures (c) and (d) of Fig.~\ref{fig_ssh}] for these OBC cases when the tight-binding hoppings include longer than nearest-neighbour terms. Furthermore, the jumps are non-existent for other models investigated --- for instance, as seen in Fig.~\ref{fig_cherny}.

Another candidate for a possible topological invariant for non-Hermitian phases is to look at vorticity, as argued in Ref.~\cite{PhysRevA.98.052116}, which we explain here in detail.
For Hermitian systems, the topological phases are characterized by the homotopy group of the space formed by the sets of eigenvectors, which can be nontrivial when there is some symmetry forcing the eigenvectors to be real. For non-Hermitian systems, however, the topological phases can be nontrivial even in the absence of symmetry --- they are described through the homotopy group of the space of gapped eigenvalues, which turns out to be the braid group $B_p$, where $p$ is the number of bands/orbitals.
The braid group is in general difficult to compute. But for two bands, since the braid group is abelian (i.e., $B_2 = \mathbb Z$), different topological phases are distinguished by integer numbers. For a closed system (i.e., with PBC), the topology of degeneracy points formed by a pair of bands, with complex energies $E_1$ and $E_2$, we define the quantity
\cite{shenzhenfu,leykambliokhhuangchongnori}
\begin{align}
\nu_{E}^{12} = \frac{1} { 2 \, \pi}
\oint_{\mathcal{C}} dk\,\partial_{k} \,\text{Arg} (E_1-E_2) \,,
\end{align}
which applies even for periodic Hamiltonians lacking a chiral symmetry \footnote{In the complex plane, we note that $\text{Arg}(z) = -i\, \ln  \big(z /|z| \big)$} and is suitable for non-Hermitian cases with complex eigenvalues.
This is often dubbed as ``vorticity invariant'' or ``eigenvalue vorticity''.
Needless to say, in systems with more than bands, the vorticity is not enough to characterize the system and it only gives the abelianization of the braid group. 
We note that the eigenvalue vorticity can only detect phase transitions which coincide with the appearance of EPs. Unlike the Zak phase discussed above, vorticity is associated with the energy dispersions of the non-Hermitian bandstructures, rather than the energy eigenstates and, hence, can detect band-touching (gap-closing) transitions. Clearly, for a Hermitian system, $\nu_{E}^{12} $ is always zero because $\text{Arg} (E_1-E_2)$ can take the values zero and one only (since $E_{1,2}$ is real) and, therefore, cannot be a nontrivial function of $ k$. More generally, $\nu_{E}^{12}$ vanishes identically whenever it involves two bands where the difference in the phase angles of their complex eigenvalues is independent of $k$.
When an EP is present in a non-Hermitian system (say at the point $k = k_{EP}$), where the two bands coalesce, we get nonzero contributions to $\nu_{E}^{12} $ and $\nu_E^{21} $ --- the vorticity is nonzero
because we have a degeneracy within a contractible loop $\mathcal{C}$ encircling the EP. 
For two bands, we can get only second-order EPs (the simplest case possible for such singular degeneracies) leading to half-integer values (or $ \mathbb Z/ 2 $)~\cite{shenzhenfu}. This can be characterized by the half-integer quantized topological invariant $\nu_{E}^{12} $ (or $\nu_E^{21} $) where $\mathcal C $ encloses a single EP.
Since the fractional values for $\nu_{E}^{12}$ and $\nu_{E}^{21}$ stem from the fact that $(E_1-E_2)$ is multi-valued and that $k_{EP}$ is a branch point in the $k$-space,
for second-order EPs, the sum $\nu_{E}= \nu_{E}^{12}  - \nu_{E}^{21} $ takes integer values and gives the eigenvalue winding number \cite{Ding2022,PhysRevA.98.052116} for the two-band case.
From these discussions and analysis, it is now clear that in the context of topological phase transitions, the calculation of vorticity does not give a complete picture as the loop enclosing a pair of EPs (with equal and opposite values of vorticity) and the loop enclosing no EPs give the same net vorticity, namely zero.


The failure of conventional topological invariants to characterize generic gap-closings in non-Hermitian models necessitates the search for a topological invariant which remains well-behaved in generic cases. We will see that the biorthogonal polarization $P$, defined below, serves this purpose as it accurately captures all topological phase transitions for open 1d chains in non-Hermitian systems.
Let us consider a one-dimensional slice of a $d$-dimensional system consisting of $N $ unit cells along this direction, harbouring $ M $ boundary states at each boundary point. 
The generalized BP operator for this slice with open boundaries is defined as \cite{flore-elisabet,flore-elisabet0}
\begin{align}
\hat P = {\mathbb{I}} - \lim_{N \rightarrow \infty} 
 \sum \limits_{ n=1}^N n\, {\hat \Pi}_n \, | \psi_{m,R} \rangle  , \quad
{\hat \Pi}_n = \sum \limits_m | e_{nm} \rangle \langle e_{nm} |\,,
\quad  | e_{nm} \rangle  = c_{nm}^\dagger |0 \rangle \,,
\end{align}
where $ {\hat \Pi}_n $ acts as the projection operator that projects a state onto the $ n^{\rm th}$ unit cell, with $m $ labeling the internal degrees of freedom inside that unit cell. The operator $c_{nm}^\dagger $ is the fermion creation operator for the fermion species labelled by $ m $ residing within the $ n^{\rm th}$ unit cell. For a 1d chain with $M$ edge modes, the $\hat P$ operator leads to the expression for the BP as
\begin{align}
 P & = 
 \text{Tr}[\mathcal{P}_{ \alpha \beta}] \,
\qquad \qquad (\text{where } 
 \mathcal{P}_{ \alpha \beta} 
 = \sum \limits_{  \alpha =1}^M
  \langle \psi_{  L, \alpha}| \, {\hat \Pi}_n \, | \psi_{ \alpha ,R} \rangle)
\nn & =
 M - \lim_{N \rightarrow \infty} 
 \sum \limits_{  \alpha =1}^M 
\frac{ \langle \psi_{ L, \alpha}| \, \sum \limits_{ n=1}^N n\, {\hat \Pi}_n \, | \psi_{ R, \alpha} \rangle
}
{\langle \psi_{ L, \alpha}|  \psi_{ R, \alpha} \rangle }\,, 
\end{align}
where $ | \psi_{ R, \alpha} \rangle \Big( | \psi_{ L, \alpha} \rangle \Big)$ is the $ \alpha^{\rm th}$ right(left) boundary eigenmode. From this expression, one can argue \cite{flore-elisabet,flore-elisabet0} that BP is a a real-space topological invariant which counts the number of boundary states localized at the boundary labelled by $ n = 1$ and, hence, it quantifies gap-closings in 1d slices with open boundaries.

\section{1d example: Non-Hermitian SSH model}
\label{sec_ssh}

This section is devoted to the study of non-Hermitian generalizations of the 1d SSH model \cite{ssh0, ssh} and Rice-Mele chains \cite{ricemele, flore_2d_chern}. A tight-binding model for such 2d chains is obtained by setting \cite{yaowang, song_yao_ssh,  han_ssh}
\begin{align}
\label{h_ssh}
d_x= t_1 + \left(t_2 +t_3 \right)  \cos k \,,\quad  
d_y = \left(t_2 - t_3 \right) \sin k +i\,\gamma/2 \,,\quad d_z = -\Delta 
\end{align}
in Eq.~\eqref{h_gen}.
The SSH version corresponds to $\Delta = 0 $.
The corresponding lattice Hamiltonian is given by
\begin{align}
\label{ssh_real}
H &= \sum \limits_n \Big[ \left( t_1+\frac{\gamma}{2} \right) 
c^\dagger_{n,A} \,c_{n,B}
+ \left( t_1 - \frac{\gamma}{2} \right) 
c^\dagger_{n,B} \,c_{n,A}
+ t_2  \left( c^\dagger_{n+1,A} \,c_{n,B} + c^\dagger_{n,B} \,c_{n+1,A} \right) 
\nn
& \hspace{ 1.2 cm}
+ t_3  \left( c^\dagger_{n,A} \,c_{n+1,B} + c^\dagger_{n+1,B} \,c_{n,A} \right)
+\Delta \left(   c^\dagger_{n,B} \,c_{n,B} - c^\dagger_{n,A} \,c_{n,A}  \right)
\Big ],
\end{align}
where $A$ and $B$ denote the two distinct sublattice sites in the bipartite lattice.

\begin{figure}[t]
	\centering
	\subfigure[]{\includegraphics[width=0.4\textwidth]{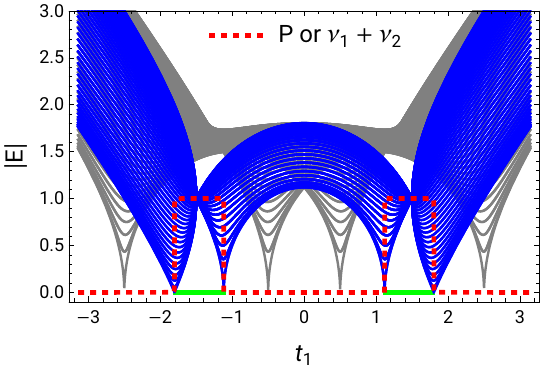}}\hspace{1 cm}
	\subfigure[]{\includegraphics[width=0.4\textwidth]{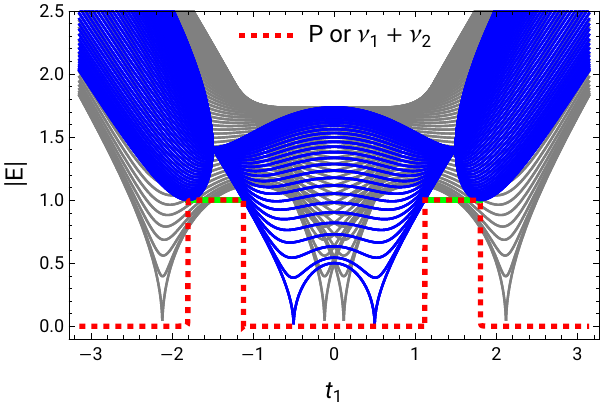}}\\
	\subfigure[]{\includegraphics[width=0.4\textwidth]{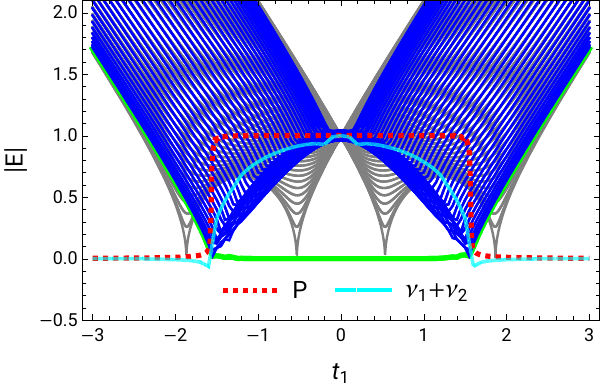}}\hspace{1 cm}
	\subfigure[]{\includegraphics[width=0.4 \textwidth]{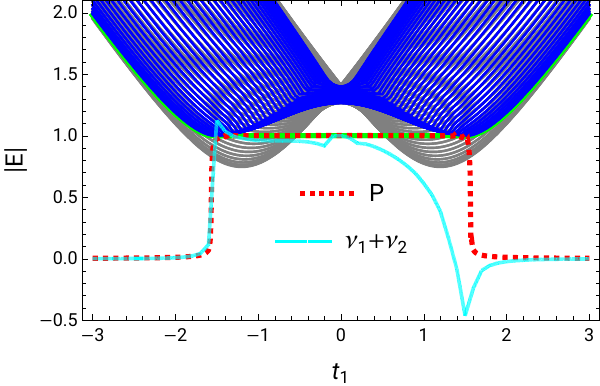}}
	\caption{\label{fig_ssh}
Absolute values of the eigenvalue spectrum for the Hamiltonian in Eq.~\eqref{ssh_real}, with the parameter values	
(a) $t_2= 1$, $t_3= 0$, $\gamma= 3$, $\Delta = 0$;	
(b) $t_2= 1$, $t_3= 0$, $\gamma= 3$, $\Delta = 1$;
(c) $t_2= 1$, $t_3=1/5$, $\gamma=4/3$, $\Delta = 0$; (d) $t_2= 1$, $t_3=1/5$, $\gamma=4/3$, $\Delta = 1$.
The spectra for the bulk states, corresponding to PBC and OBC cases, have been shown in gray and blue, respectively. The edge mode energies have been indicated with green. The biorthogonal polarization $P$ (depicted via the dashed red lines) jumps precisely by unity when the edge enters or leaves the bulk spectrum (blue), and these jumps have no correspondence with the location of the EPs of the periodic spectrum.
We have also shown the total non-Bloch band theory winding $\nu_{tot}$ (demarcated by cyan curves) for the two bands. In (a) and (b), 
$\nu_{tot}$ coincides with $P$. But in (c) and (d), $\nu_{tot}$ deviates from a quantized value.
Nevertheless, it still shows jumps at the phase transitions points captured by $P$. Denoting the number of unit cells taken for plotting the OBC spectra and computing $P$ by $ N_\textrm{OBC}$ and $ N_P$, respectively, we have used $\lbrace N_\textrm{OBC}, N_P \rbrace $ equal to (a) $\lbrace 100, 3500 \rbrace $; (b) $\lbrace 100, 3000 \rbrace $; (c) $\lbrace 45, 45 \rbrace $; and (d) $\lbrace 45, 45 \rbrace $.
}
\end{figure}

The Hamiltonian has a chiral symmetry when $d_z=0$. For an open chain, using the GBZ, we compute the $ \nu_{tot} = \nu_1 +\nu_2 $, which should be an integer (as only 2EPs are possible for this two-band case). We find that both $\nu_{tot}$ and biorthogonal polarization $P$ identify the phase transition points via sharp jumps, at the points where the edge modes enter into or separate out from the bulk modes. However, for the non-chiral case, $\nu_{tot}$ is not quantized in all cases [cf. Fig.~\ref{fig_ssh}(d)].

For the $t_3 = 0 $ cases, Eq.~\eqref{eq_beta2} is quadratic in  $\beta$, and can 
be solved easily in closed forms. The allowed solutions are the ones when the two roots are equal in magnitude. In fact, we find that the solutions have $|\beta| = {\Gamma}$, with $\Gamma =\sqrt{   |2\,t_1 -\gamma|  /{| 2\,t_1 + \gamma|} }$. Since $\beta$ has a constant magnitude, the GBZ is a circle with radius $\Gamma$. Hence, the simple substitution $k  = \theta - i\ln \Gamma $ (with $ \theta \in [0, 2\pi)$) in the periodic Hamiltonian $H(k)$, leading to $H(\beta = \Gamma \,e^{i\,\theta})$, gives us the energy spectrum (as well as the eigenvectors) for the open system, as outlined in Refs.~\cite{flore-elisabet0,flore-elisabet}.
We have used the analytical expressions derived in Refs.~\cite{flore-elisabet0,flore-elisabet} to compute $P$. For Fig.~\ref{fig_ssh}(a) and (b), $N=3500$ and $N = 3000$ have been used for the summations involved in the expression for $P$.

The nonzero $t_3$ case corresponds to longer-ranged hoppings (in the sense of the $A$-$B$ sublattice-site-arrangements of Fig.~1 of Ref.~\cite{yaowang} and Fig.~2 of Ref.~\cite{han_ssh}) and, hence, the open chain does not admit analytical expressions for the energy eigenvalues, eigenvectors, or BP. In fact, the polynomials in Eqs.~\eqref{eq_beta} and \eqref{eq_beta2} are fourth order in $\beta$, and the admissible solutions are the ones when the roots satisfy $|\beta_2| =|\beta_3|$ (after ordering the $\beta_j$'s according to their magnitude). This gives us the GBZ, and we find that it forms a closed loop in the complex $\beta$-plane where the magnitude of $\beta$ also changes. In other words, unlike the $t_3=0$ case, the GBZ is no longer a circle. The Hamiltonian $H(\beta)$, with the admissble values of $\beta$ fed in, give us the energy and eigenvctors of the open system. Alternatively, we can of course determine all these features by considering the Hamiltonian of the open chain in the real space, with $N $ unit cells.
In Fig.~\ref{fig_ssh}(c) and (d), we have computed the spectrum and $P$ by considering the real-space Hamiltonian of length $N=45 $. On the other hand, the non-Bloch or generalized winding vector has been calculated by computing the GBZ.

\section{2d example: Non-Hermitian Chern insulator}
\label{secchern}

In this section, we study 2d models built by stacking Rice-Mele chains \cite{ricemele, flore_2d_chern} and generalized to harbour non-Hermitian hopping terms.\footnote{A different non-Hermitian version of the Rice-Mele chain has been studied in Ref.~\cite{wangzhangsong}} This model exhibits a non-Hermitian analogue of a Chern insulator phase such that chiral edge states exist in the phase space. The Bloch Hamiltonian can be obtained from Eq.~\eqref{h_gen} by setting \cite{flore-elisabet}
\begin{align}
\label{eq_chern}
d_x ({\mathbf k}) &= t_+ (k_y) 
+ \left [\,   t_-(k_y) + t_3 \,\right ] \cos k_x\,,\quad
d_y ({\mathbf k}) = 
\left[  \,t_-(k_y) - t_3 \,\right ] \sin k_x +  i\,\gamma / 2\,, 
\quad d_z ({\bf k}) =  - \,\Delta \, \sin k_y \,,\nn
t_\pm (k_y) &= t_1 \pm \delta \, \cos k_y\,.
\end{align} 
Here we consider an open system with OBC either along the $x$-axis or the $y$-axis, while retaining PBC along the other axis.

\subsection{OBC along the $x$-axis}

For OBC along the $x$-axis, the generalized Bloch Hamiltonian obtained by
\begin{align}
\lbrace H(k_x,k_y), \, k_x \in \mathds{R} \rbrace \rightarrow 
\lbrace H(\beta=e^{i\,k_x}, k_y), \, k_x \in \mathds{C} \rbrace \,.
\end{align}
For a given $k_y$, we have a system similar to the 1d models considered in Sec.~\ref{sec_ssh}.

\subsubsection{Model for $t_3 =0 $ admitting closed-form analytical solutions}
\label{secsolvable}

As seen in Sec.~\ref{sec_ssh}, the $t_3 = 0 $ can be solved in terms of closed-form expressions and the GBZ for the effective open 1d system for a given $k_y$ is a circle. Since the $\beta$ of the GBZ has a constant magnitude, the open system can be characterized simply by incorporating the shift
$e^{i\,k_x} = e^{i\,\theta}\, e^{ \ln \Gamma} $ in the PBC Hamiltonian, where 
\begin{align}
\eta_x(k_y)  =\frac{t_+(k_y)} {t_-(k_y)}\,,\quad
\eta_y(k_y) =\frac{i\,\gamma} {2\,t_-(k_y)}\,,\quad
\eta_z(k_y) = -\frac{\Delta \sin k_y} {t_-(k_y)}\,, \quad
\Gamma
=\sqrt{ \left| \frac{\eta_x + i\,\eta_y } 
{\eta_x - i \,\eta_y} \right |} 
=\sqrt{ \left | \frac{2\,t_+  -   \gamma} {2\,t_+ + \gamma} \right | }\,.
\end{align}
Explicitly, the generalized Bloch Hamiltonian for the open system takes the form
\begin{align}
\label{eqhamobc}
H_{\textrm{OBC}}(k_y,\theta) & =t_- \left[ \,\mathcal{H}_0(\theta) 
+ \boldsymbol \eta(k_y) \cdot \boldsymbol{\sigma} \,\right] ,\quad
\mathcal{H}_0(\theta)  
=\begin{pmatrix}
0 & e^{-i\,\theta }\\
e^{i\,\theta} & 0\\
\end{pmatrix} ,
\end{align}
with the energy eigenvalues $\pm E_\textrm{OBC}$, where
\begin{align}
E_\textrm{OBC} & = 
\sqrt{ t_+^2 + t_-^2 -{\gamma ^2} / {4}
+ \Delta ^2 \sin ^2 k_y 
+  t_- \left[ \,
\Gamma  \,e^{i \,\theta } 
\left(  t_+ + \gamma / {2}  \right)
+ e^{-i \,\theta } \left(t_+- {\gamma } /{2}\right) /{\sqrt{\Gamma }}
\, \right]  }\,.
\end{align}

For the sake of completeness, we first determine the bulk states, using the techniques described in Refs.~\cite{yaowang,yaosongwang,flore-vatsal}. 
The state in the $n^{\text{th}}$ unit cell of the open system is given by
\begin{align}
	\tilde{\Psi}_{R,\pm}(k_y,\theta,n) = 
\Gamma^{n}\, e^{i\,n\,\theta }\,\tilde{\psi}_{R,\pm}(k_y,\theta)\,,
\end{align}
with
\begin{align}
\tilde{\psi}_{R,\pm}(k_y,\theta) & = \begin{pmatrix}
-\Delta \sin k_y \pm E_\textrm{OBC}(k_y,\theta)
\\ 
t_1 +\delta  \cos  k_y 
-{\gamma }/2
+ \left(t_1-\delta  \cos  k_y \right) \Gamma\,  e^{i \,\theta}
\end{pmatrix} .
\end{align}
The bulk state at the $n^{\text{th}}$ unit cell can be written as a superposition of $\tilde{\Psi}_{R,\pm}(k_y,\theta,n)$ and $\tilde{\Psi}_{R,\pm}(-k_y,-\theta,n)$ [with $-\theta$ being equivalent to $\left( 2\pi-\theta \right) $] as follows:
\begin{align}
\Psi_{R,\mathrm{Bulk},\pm,\alpha}(k_y,\theta,n) 
& = C_1  \, \tilde{\Psi}_{R,\pm,\alpha}(k_y,\theta,n) 
+ C_2\, \tilde{\Psi}_{R,\pm,\alpha}(-k_y,  -\theta,n)
\nn &  =  \Gamma^{n}
\left[C_1 \,e^{ i\, n\,\theta } \,\tilde{\psi}_{R,\pm,\alpha} (k_y,\theta)
+ C_2 \,e^{-i\,n\,\theta }
\,\tilde{\psi}_{R,\pm,\alpha}(-k_y,-\theta)\right],   
\end{align}
The label $\alpha \in \lbrace A, B\rbrace $ in $\Psi_{R,\mathrm{Bulk},\pm,\alpha}(k_y,\theta,n)$
refers to the amplitude of the wavefunction on sublattice $\alpha$.
Furthermore, $\theta = \pi \,j /N$ with $j \in [ 1,N-1]$, and the boundary condition 
\begin{align}
	\Psi_{R,\mathrm{Bulk},\pm,B}(k_y,\theta,0)
= \Psi_{R,\mathrm{Bulk},\pm,B}(k_y,\theta,N) = 0
\end{align}
has to be imposed.
The boundary condition leads to
\begin{align}
&	\frac{C_2}  {C_1} 
= -\frac{\tilde{\psi}_{R,\pm,B}(\mathbf k)}{\tilde{\psi}_{R,\pm,B}(-\mathbf k)} = -1
 \Rightarrow
\Psi_{R,\mathrm{Bulk},\pm,\alpha}(k_y,\theta,n) 
= \Gamma^{n}
\left[e^{i\, n\,\theta}  \,\tilde{\psi}_{R,\pm,\alpha}(k_y,\theta)
-e^{-i \,n\,\theta} \,
\tilde{\psi}_{R,\pm,\alpha}(-k_y,-\theta)\right].   
\end{align}
Now we can make an educated ansatz for the (unnormalized) bulk states as
\begin{align}
\label{eq_bulk_states}
	\Psi_{R,\mathrm{Bulk},\pm} (k_y,\theta) = \begin{pmatrix}
	\Psi_{R,\mathrm{Bulk},\pm,A} (k_y,\theta,1) \\
	\Psi_{R,\mathrm{Bulk},\pm,B} (k_y,\theta,1)\\
	\Psi_{R,\mathrm{Bulk},\pm,A}(k_y,\theta,2)\\
	\Psi_{R,\mathrm{Bulk},\pm,B}(k_y,\theta,2)\\
	\vdots\\
	\Psi_{R,\mathrm{Bulk},\pm,B}(k_y,\theta,N-1)\\
	\Psi_{R,\mathrm{Bulk},\pm,A}(k_y,\theta,N)\\
	\end{pmatrix}.
\end{align}
A straightforward application of the eigenvalue condition
$H_{\textrm{OBC}} \,
\Psi_{R,\mathrm{Bulk},\pm} = \pm E_\textrm{OBC} 
\, \Psi_{R,\mathrm{Bulk},\pm} $ shows that this indeed represents a right eigenstate with energy $ \pm E_{\textrm{OBC}} (k_y,\theta)$. By making use of the fact $H_{\textrm{OBC}} ^\dagger$ is simply obtained by transforming $\gamma \rightarrow - \gamma$ in the Hamiltonian $ H_{\textrm{OBC}} $, the left eigenstates are readily figured out by taking the complex conjugation of the right eigenstates and by setting $\gamma \rightarrow - \gamma$. 

\begin{figure}[t]
	\centering
	\subfigure[]{\includegraphics[width=0.4\textwidth]{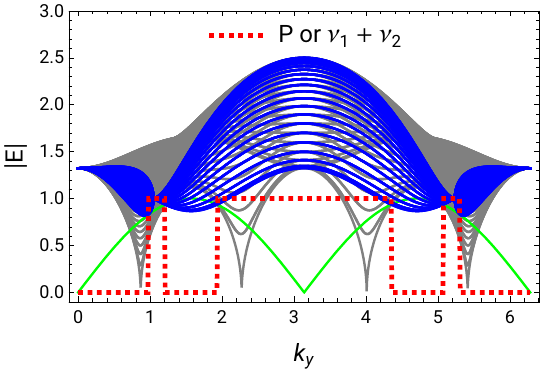}} \hspace{1 cm}
	\subfigure[]{\includegraphics[width=0.4 \textwidth]{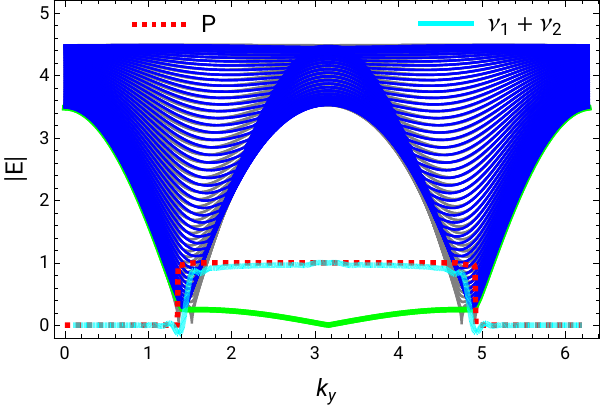}}\\
	\caption{\label{fig_chernx}
Absolute values of the energy spectrum for the Hamiltonian $ { \mathbf{d}} ({\mathbf k})  \cdot \boldsymbol \sigma$, with the components of ${ \mathbf{d}}$ given by Eq.~\eqref{eq_chern}, after imposing OBC along the $x$-axis. The chosen parameter values are
(a) $t_1= \delta=\Delta= 1$, $t_3 =0 $, $\gamma=3$ (cf. the model of Sec.~\ref{secsolvable} admitting closed-form analytical solutions) and	
(b) $t_1= \delta=2$, $t_3=1/2$, $\gamma= 4/5$, $\Delta = 1/4$.
The spectra for the bulk states, corresponding to the PBC and OBC cases, have been shown in gray and blue, respectively. The edge mode energies have been indicated with green. The biorthogonal polarization $P$ (depicted via the dashed red lines) jumps precisely by unity when the edge enters or leaves the bulk spectrum (blue), and these jumps have no correspondence with the location of the EPs of the periodic spectrum.
We have also shown the total non-Bloch band theory winding $\nu_{tot}$ (demarcated by cyan curves) for the two bands. In (a), $\nu_{tot}$ coincides with $P$. But in (b), $\nu_{tot}$ deviates from a quantized value, as for non-chiral systems, such a topological number does not always take quantized values.
Nevertheless, it still shows jumps at the phase transitions points captured by $P$.
Denoting the number of unit cells taken for plotting the OBC spectra and computing $P$ by $ N_\textrm{OBC}$ and $ N_P$, respectively, we have used $\lbrace N_\textrm{OBC}, N_P \rbrace $ equal to (a) $\lbrace 80, 3000 \rbrace $ and (b) $\lbrace 50, 500 \rbrace $.
}
\end{figure}

For OBCs with a broken unit cell terminating with an $A$  sublattice site at each end of the 1d chain, we find that the wavevectors \cite{flore-emil,flore-elisabet,flore-elisabet0}
\begin{align}
| \psi_{ R} \rangle  = \mathcal{N}_R \sum \limits_{n=1}^{N} r_R^n \,
 c_{n, A}^\dagger \, | 0\rangle \text{ and }
| \psi_{L}  \rangle  = \mathcal{N}_L \sum \limits_{n=1}^{N} r_L^n \,
 c_{n, A}^\dagger \,| 0\rangle
\end{align}
characterize a mode with eigenvalue $d_z$, with $\mathcal{N}_R $ and $\mathcal{N}_L $ representing the corresponding normalization factors. Here
\begin{align}
r_R = -\frac{ t_1 + \delta  \cos k_y  -{\gamma } / {2}} 
{t_1-\delta  \cos k_y } \,, \quad
r_L = -\frac{ t_1 
+ \delta  \cos  k_y  + {\gamma } / {2} }
{ t_1-\delta  \cos k_y }\,,
\end{align}
and $c_{n, A}^\dagger$ is the fermion creation operator at the $A$ sublattice site for the $n^{\rm th}$ unit cell. This mode has the following characteristics:
\begin{enumerate}

\item For $ |r_L^* \, r_R | <1 $, it is exponentially localized at
the unit cell $n=1$.

\item For $ |r_L^* \, r_R | > 1 $, it is exponentially localized at
the unit cell $n= N $.

\item For $ |r_L^* \, r_R | = 1 $, it is delocalized and forms/merges with the bulk states.
 
\end{enumerate}
Hence, the last condition marks the singular point(s) at which the number and/or
localization properties of boundary modes change.
Using the explicit expressions of all the eigenstates shown above, we can compute the BP for a 1d slice at a given $k_y$-value.

\begin{figure}[t]
	\centering
	\subfigure[]{\includegraphics[width=0.27 \textwidth]{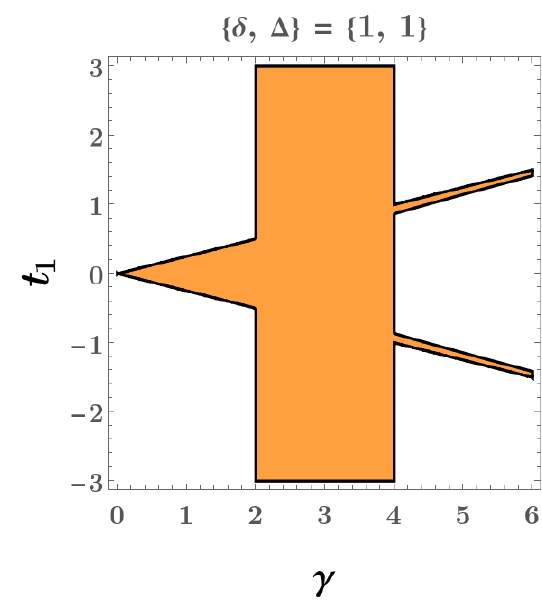}} \hspace{1 cm}
	\subfigure[]{\includegraphics[width=0.27 \textwidth]{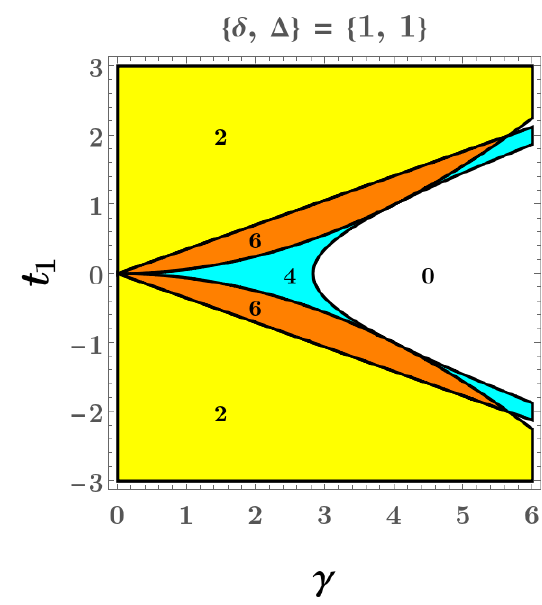}}\\
	\caption{\label{fig_phases}
Bulk phase diagrams for the Hamiltonian  $ { \mathbf{d}} ({\mathbf k})  \cdot \boldsymbol \sigma$, with the components of ${ \mathbf{d}}$ given by Eq.~\eqref{eq_chern}, against the $\gamma$-$t_1$ plane. Using the parameter values $  \delta=\Delta= 1$ and $t_3 =0 $, this gives us the model of Sec.~\ref{secsolvable}, which allows closed-form analytical expressions. The two subfigures represent the PBC and OBC cases, respectively. The bulk eigenstate-spectra of Fig.~\ref{fig_chernx}(a) are related to the phases shown here. For subfigure (a), the orange region corresponds to one which harbours EP solutions, whereas the white region represents the regime where no EP solution exists. The different regions in subfigure (b) are distinguished by the number of distinct admissible solutions for $k_y$ for which an edge mode becomes delocalized, merging with the bulk, as given by Eq.~\eqref{eqepsolns}. Those numbers are also indicated explicitly against the corresponding colour-coded regions. 
}
\end{figure}

Now we consider the possible topological phase transitions in the effective 1d systems, which arise when a chiral edge mode enters or leaves the bulk modes.
In the bulk of the PBC spectrum, EPs appear at (1) $ \sin  k_y 
= \pm \sqrt{\frac{{\gamma ^2} / {4}-4 \,t_1^2} {\Delta ^2}} $ for $k_x=0$, and
(2) $\sin  k_y 
=  \pm \sqrt{\frac{ {\gamma ^2} / {4}-4\, \delta ^2}{\Delta ^2-4 \,\delta ^2}} $ for $k_x= \pi $.
In the OBC spectrum, gaps close at points given by
\begin{align}
\label{eqepsolns}
\cos k_y = \frac{\gamma ^2} {16 \,\delta \, t_1} \text{ and }
\cos k_y = \pm \,\frac{\sqrt{\gamma ^2/8 - t_1^2}} { \delta } \,,
\end{align}
which are derived from the condition $ |r_L^* \, r_R | = 1 $ described above.
Thus, it is apparent that we have regions admitting zero, two, four, and six values of $k_y$ where the chiral band merges with the bulk states.
Since these points in the parameter space as those where the chiral mode attaches to/detaches from the bulk bands \cite{flore-elisabet}, the value of BP jumps by $\pm 1 $ while crossing them.
In Fig.~\ref{fig_chernx}(a), we show the spectrum, with the singular points identified by the behaviour of $P$ and $\nu_{tot}$, which corroborate the analytical arguments and expressions presented so far.
Fig.~\eqref{fig_phases} illustrates the phase diagram for the spectra of the bulk eigenstates both for the PBC and OBC cases. 

\subsubsection{Model for $t_3 \neq 0 $}
\label{secnumerical}

The nonzero $t_3$ case corresponds to longer-ranged hoppings (when we arrange the sublattice sites as shown in Fig.~1 of Ref.~\cite{yaowang} and Fig.~2 of Ref.~\cite{han_ssh}), and hence the open chain does not admit analytical expressions for the energy eigenvalues, eigenvectors, or BP. Hence, we have to fund the physical quantitites numerically.
In Fig.~\ref{fig_chernx}(b), we show the eigenvalue spectrum along with the behaviour of $P$ and $\nu_{tot}$. Similar to the SSH model studied in Sec.~\ref{sec_ssh}, we find that although $\nu_{tot}$ shows jumps at the phase transition points where an edge modes enters/leaves the bulk states, its value does not remain quatized within a given phase. On the other hand, BP shows an unambiguous quantized behavior with appropriate jumps at the phase transition points.

\subsection{OBC along the $y$-axis}
 
\begin{figure}[t]
	\centering
	\subfigure[]{\includegraphics[width=0.4\textwidth]{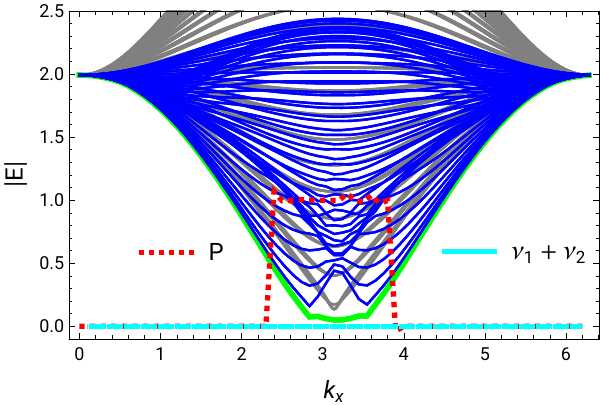}} \hspace{1 cm}
	\subfigure[]{\includegraphics[width=0.4 \textwidth]{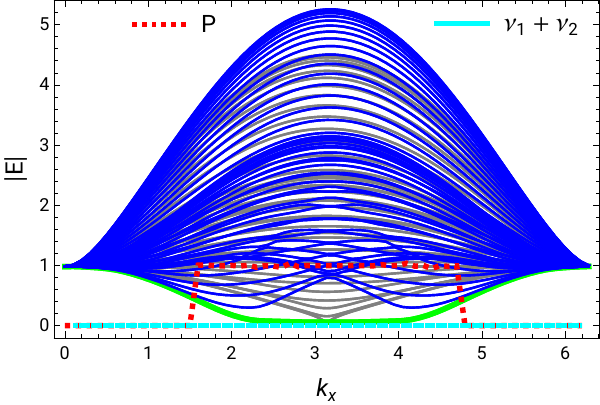}}\\
\caption{\label{fig_cherny}
Absolute values of the energy spectrum for the Hamiltonian $ { \mathbf{d}} ({\mathbf k})  \cdot \boldsymbol \sigma$, with the components of ${ \mathbf{d}}$ given by (a) Eq.~\eqref{eq_chern2} and
(b) Eq.~\eqref{eq_chern3}, after imposing OBC along the $y$-axis. For both the subfigures, the parameter values have been set to $t_1=1$, $ \delta= 1.75 $, $t_3 =0 $, $\gamma=2/5$, and $\Delta = 0.1 $.
The spectra for the bulk states, corresponding to the PBC and OBC cases, have been shown in gray and blue, respectively. The edge mode energies have been indicated with green. The biorthogonal polarization $P$ (depicted via the dashed red lines) jumps precisely by unity when the edge enters or leaves the bulk spectrum (blue), and these jumps have no correspondence with the location of the EPs of the periodic spectrum.
We have also shown the total non-Bloch band theory winding $\nu_{tot}$ (demarcated by cyan curves) for the two bands. In both cases, $\nu_{tot}$ does not show any jump at the phase transitions points captured by $P$.
Denoting the number of unit cells taken for plotting the OBC spectra and computing $P$ by $ N_\textrm{OBC}$ and $ N_P$, respectively, we have used $\lbrace N_\textrm{OBC}, N_P \rbrace = \lbrace 50, 600 \rbrace $ for both the subfigures.
}
\end{figure}

For OBC along the $y$-axis, the generalized Bloch Hamiltonian obtained by
\begin{align}
\label{eq_cherny_ham}
\lbrace H(k_x,k_y), \, k_y \in \mathds{R} \rbrace \rightarrow 
\lbrace H(k_x, \beta=e^{i\,k_y}), \, k_y \in \mathds{C} \rbrace \,.
\end{align}
For a given $k_y$, we have a system similar to the 1d models considered in Sec.~\ref{sec_ssh}. 

Expanding the components of $\mathbf d$, we see that the choice
\begin{align}
\label{eq_chern2}
d_x ({\mathbf k}) &=  t_1 \,(1+ \cos k_x)
+ \delta \,(1- \cos k_x) \cos k_y\,,\quad
d_y ({\mathbf k}) = 
( t_1 \sin k_x +  i\,\gamma / 2)- \delta \sin k_x \cos k_y   \,, 
\quad d_z ({\bf k}) =  - \Delta \, \sin k_y \,,
\end{align} 
leads to the lattice hoppings
\begin{align}
\label{chern_lat}
&t_1 \,(1+ \cos k_x) \left[ c^\dagger_{j_y,A}\,c_{j_y,B}
+ c^\dagger_{j_y,B}\,c_{j_y,A} \right ]
+ \frac{ \delta \,(1- \cos k_x) } {2}
\left[  c^\dagger_{j_y,A}\,c_{j_y+1,B}
+ c^\dagger_{j_y,B}\,c_{j_y+1,A}
+ c^\dagger_{j_y+1,A}\,c_{j_y,B}
+ c^\dagger_{j_y+1,B}\,c_{j_y,A} 
\right] \nn
& + ( \gamma / 2 -i \,t_1 \sin k_x )
\left[ c^\dagger_{j_y,A}\,c_{j_y,B} -  c^\dagger_{j_y,B}\,c_{j_y,A} \right ]
+\frac{i\, \delta \sin k_x }  {2}
\left[ c^\dagger_{j_y,A}\,c_{j_y+1,B}- c^\dagger_{j_y,B}\,c_{j_y+1,A}
+ c^\dagger_{j_y+1,A}\,c_{j_y,B}-  c^\dagger_{j_y+1,B}\,c_{j_y,A} 
\right] \nn
& +\frac{ i\,\Delta } {2} \left[
c^\dagger_{j_y,A}\,c_{j_y+1,A}-c^\dagger_{j_y,B}\,c_{j_y+1,B}
- c^\dagger_{j_y+1,A}\,c_{j_y,A}+ c^\dagger_{j_y+1,B}\,c_{j_y,B}
\right ] .
\end{align}
In this case, Eq.~\eqref{eq_beta2} takes the form:
\begin{align}
 \Big[ &
 \delta  \left \lbrace 
 \delta +e^{i\, \phi }
 \left(\beta ^2 \,\delta  \left(1 + e^{i \, \phi }\right)-\beta \, \gamma 
 +\delta \right)\right \rbrace
+ \left(1+e^{i \,\phi }\right) 
 \left(1+\beta ^2 \,e^{i \,\phi }\right) \left( \Delta ^2 - 2\, \delta ^2  \right)
  e^{i \,k_x}
 \nn &   +\delta \, e^{2 \,i \,k_x} \left \lbrace \delta + e^{i\, \phi } 
  \left( \delta +\beta  \left(\gamma +\beta \, \delta  \left(1 + e^{i \,\phi }\right)\right)
  \right) \right \rbrace  \Big ]
\left(1 - e^{-i \,\phi } \right) \left(\beta ^2 - e^{-i \,\phi }\right) 
e^{-i \,k_x} =0 \,.
\end{align}
The procedure to obtain the admissible $\beta$-values gives us only the value unity for $|\beta|$, which actually comes from the factor $ \left(\beta ^2 - e^{-i \,\phi }\right) $ [giving $\beta = \pm  \,e^{-i \,\phi /2} $] on the left-hand-side of the equation.

We also consider a slightly modified model with
\begin{align}
\label{eq_chern3}
d_x ({\mathbf k}) &=  t_1 \,(1+ \cos k_x)
+ \delta \,(1- \cos k_x) \cos k_y\,,\quad
d_y ({\mathbf k}) =   i\,\gamma / 2- \delta \, \cos k_y   \,, 
\quad d_z ({\bf k}) =  - \Delta \, \sin k_y \,.
\end{align} 
For the OBC along the $y$-axis, Eq.~\eqref{eq_beta2} now takes the form:
\begin{align}
\Big [ &
e^{i \, \phi } \left\{\left(\beta ^2+1\right) \Delta ^2
+2 \,i \,\beta \, \gamma \, \delta  \sin  k_x 
-2 \,\delta  \left(\beta ^2 \,\delta  +\delta 
+2 \,\beta \, t_1   \right) \right\}
+2 \,\delta  \cos k_x 
\left\{\delta +e^{i \, \phi } \left(\beta ^2 \,\delta  
\left(1+e^{i \, \phi }\right)+\delta +2 \,\beta \, t_1\right)\right\}
\nn &  + \beta ^2 \,e^{2\, i \,\phi } 
\left( \Delta ^2 - 2 \,\delta ^2  \right)-2\, \delta ^2+\Delta ^2
\Big] 
\left(1-e^{-i \,\phi }\right) \left(\beta ^2-e^{-i \,\phi }\right)=0\,.
\end{align}
Here also the admissible $\beta$-values give us only the value unity for $|\beta|$, which again comes from the factor $ \left(\beta ^2 - e^{-i \,\phi }\right) $ on the left-hand-side of the equation.

For both these models, for OBC along the $y$-axis, the GBZ is found to coincide with the normal BZ, i.e., $\beta = e^{i\,\theta}$ with the magnitude always taking the value unity, thus reproducing the PBC spectrum (rather than the OBC spectrum). This means the GBZ scheme simply fails to capture the correct physics for these cases, failing to even capture the correct energy spectrum.
Now we consider the possible topological phase transitions in these effective 1d systems, which arise when a chiral edge mode enters or leaves the bulk modes. In Fig.~\ref{fig_cherny}, we show the spectrum, supplemented by the behaviour of $P$ and $\nu_{tot}$. We find that whereas $P$ captures the phase transitions when the edge modes enter or leave the bulk modes, $\nu_{tot}$ remains zero throughout.

\section{Summary and outlook}
\label{secsum}

In this paper, we have studied open 1d chains in 1d and 2d systems with non-Hermitian couplings, which are captured by tight-binding models with two bands. Our aim has been to characterize topological phases separated by edge modes leaving/merging with the bulk modes. From our investigations, we have found that conventional quantities like winding numbers, which work well for Hermitian systems, fail for generic open non-Hermitian systems, as they do not possess a distinct quantized value within a given phase. However, we have identified the biorthogonal polarization, a real-space topological invariant, to correctly capture the aforementioned phase transitions as it retains the quantized value of one or zero, depending on whether a given edge mode exists or whether it loses its edge mode character by getting delocalized and absorbed within the bulk.

BP can characterize gap-closings only for 1d slices of a non-Hermitian system of generic dimensionality. Hence, in order to consider OBCs along more than one direction, it will be worthwhile to search for an analogue of the BP which can characterize gap-closings for edges of arbitrary codimension. This will serve, for example, as Chern numbers for higher-dimensional non-Hermitian matrices with open boundary conditions. We would like to emphasize that all this is necessary because of the crucial fact that the PBC spectra and OBC spectra are completely different for non-Hermitian systems, with no familiar notions of bulk-boundary which exist for Hermitian situations \cite{slager,Okuma2020,emil_review}. 
We would like to add that another quantity that has been discussed in the literature is the \textit{Electronic Polarization} $ z^{(1)} $ \cite{nakamura1,nakamura2}. We leave a closer investigation of its effectiveness for future work.

\section*{Acknowledgments}
We thank Emil J. Bergholtz for suggesting the problem. We are also grateful to Elisabet Edvardsson and Kang Yang for useful discussions. This research, leading to the results reported, has received funding from the European Union's Horizon 2020 research and innovation programme under the Marie Skłodowska-Curie grant agreement number 754340.

\bibliography{ref}
\end{document}